\documentclass[a4paper, 10pt, conference]{ieeeconf}      
\IEEEoverridecommandlockouts	

\usepackage{multirow}
\usepackage{lipsum}
\usepackage{amsmath}
\usepackage{amssymb}
\usepackage{subcaption}
\usepackage{graphicx}
\usepackage{diagbox}
\graphicspath{{./Figures/}}
\usepackage{glossaries}
\usepackage[table,xcdraw]{xcolor}
\usepackage[ruled,vlined]{algorithm2e}
\usepackage[final]{listings}
\usepackage{longtable}

\usepackage[hidelinks]{hyperref}
\usepackage{tikz,pgfplots}

\newacronym{ITS}{ITS}{Intelligent Transportation Systems}
\newacronym{ROS}{ROS}{Robot Operating System}
\newacronym{URDF}{URDF}{Unified Robot Description Format}
\newacronym{SLAM}{SLAM}{Simultenous Localization and Mapping}
\newacronym{JSON}{JSON}{JavaScript Object Notation}
\newacronym{LDS}{LDS}{Laser Distance Sensor}
\newacronym{iCab}{iCab}{Intelligent Campus Automobile}
\newacronym{AV}{AV}{Autonomous Vehicle}
\newacronym{HMI}{HMI}{Human Machine Interface}
\newacronym{PAFs}{PAFs}{Part Affinity Fields}
\newacronym{VRUs}{VRUs}{Vulnerable Road Users}

\title{\LARGE \bf Response of Vulnerable Road Users to Visual Information from Autonomous Vehicles in Shared Spaces}

\author{Walter Morales Alvarez$^{1}$ \emph{Student Member, IEEE}, Miguel  \'Angel de Miguel$^2$, Fernando Garc\'ia$^2$ \emph{Member, IEEE}\\ and Cristina Olaverri-Monreal$^{1}$ \emph{Senior Member, IEEE}%
\thanks{$^1$ Johannes Kepler University Linz, Austria; Chair for Sustainable Transport Logistics 4.0. \texttt{\{walter.morales\_alvarez, cristina.olaverri-monreal\}@jku.at}}%
\thanks{$^2$ Universidad Carlos III de Madrid, Spain.
	\texttt{\{mimiguel, fegarcia\}@ing.uc3m.es}}%
}

\newcommand\copyrighttext{%
  \footnotesize \textcopyright 2019 IEEE. Personal use of this material is permitted. Permission from IEEE must be obtained for all other uses, in any current or future   media, including reprinting/republishing this material for advertising or promotional purposes, creating new collective works, for resale or redistribution to servers or lists, or reuse of any copyrighted component of this work in other works.  DOI: \href{https://ieeexplore.ieee.org/document/8917501}{10.1109/ITSC.2019.8917501} }

\newcommand\copyrightnotice{%
\begin{tikzpicture}[remember picture,overlay]
\node[anchor=south,yshift=10pt] at (current page.south) {\fbox{\parbox{\dimexpr\textwidth-\fboxsep-\fboxrule\relax}{\copyrighttext}}};
\end{tikzpicture}%
}

\begin{document}

\maketitle
\copyrightnotice
\thispagestyle{empty}
\pagestyle{empty}

\begin{abstract}
Completely unmanned autonomous vehicles have been
anticipated for a while. Initially, these are expected to drive only under certain conditions on some roads, and advanced functionality is required to cope with the ever-increasing challenges of safety. To enhance the public's perception of road safety and trust in new vehicular technologies, we investigate in this paper the effect of several interaction paradigms with vulnerable road users by developing and applying algorithms for the automatic analysis of pedestrian body language. We assess behavioral patterns and determine the impact of the coexistence of AVs and other road users on general road safety in a shared space for VRUs and vehicles.
Results showed that the implementation of visual communication cues for interacting with VRUs is not necessarily required for a shared space in which informal traffic rules apply. \\

\end{abstract}


\section{Introduction}
\label{sec:introduction}

The arrival of driverless vehicles has been anticipated already for some time. Several aspects regarding their convenience and safety have been addressed, highlighting for example that the replacement of the human driver by automation will lead to more efficient driving patterns that result in the environmental benefits of decreased traffic congestion, as well as public safety improvements due to fewer traffic-related injuries~\cite{olaverri2016autonomous}. Many vehicles are already equipped with the technology that enable self-driving automation, such as lane-keeping assistance and automated braking. 
In the near future highly autonomous, complex dynamical systems will be mature enough to implement intelligent autonomous vehicles (AV). \\

Competition has been generated between the different automotive industries to develop their autonomous vehicles at ever higher levels of automation and to commercialize them. BMW showed an autonomous concept car at CES 2016 and announced its initiative to include the automation of their vehicles as part of their iNEXT project \cite{websitebmw}. General Motors launched in 2018 the Cadillac CT6, its first autonomous vehicle level 2, which possesses a hands-free driving system called Super Cruise \cite{websitecompaniesworkingav}. 
At the same time, Renault developed its own autonomous concept vehicle, the EZ-GO presented at the Geneva Motor Show in 2018 \cite{websiterenault}. They also announced that in 2020 they would present a fleet of vehicles equipped with a significant amount of autonomous functionality. Finally, Nissan introduced its Serena model, boasting a one-lane autonomous driving system called proPILOT and which is being featured in Japan~\cite{websitenissan}.
Despite the multiple advances that have taken place all over the world that bring society closer to self-driving vehicles, there is still not a car on the road that is completely autonomous. Most are concepts and prototypes of fully automated vehicles that drive in controlled environments (e.g.: universities) under certain conditions and on predetermined roads. 

Road safety is not only determined by the technology of the autonomous vehicles themselves, but rather a significant aspect of safety lies in the interaction between the automated vehicles and the other vulnerable road users (VRUs). 

Therefore it is crucial to assess patterns regarding complexity and risk by judging and anticipating the actions of the different actors in the system to determine the rules for their co-existence. In this context perceived trustworthiness of new vehicular technologies will be jeopardized if other users are not able to determine the authenticity of the information provided by the autonomous vehicle~\cite{allamehzadeh2016automatic}. 

In this paper we aim at increasing road safety through approaches that augment awareness of the surrounding environment for road users and the automation. 
We focus on VRUs,  as they cannot make visual contact with a driver in a driverless vehicle and they must therefore turn to novel or unfamiliar ways of understanding the decisions made by the vehicle, if they are able to do so at all.

To address this, we investigate interaction strategies by applying in field tests the algorithms for the automatic analysis of pedestrian body language presented in~\cite{Morales-Alvarez2018} to determine the impact of the coexistence of AVs and other road users on general road safety. 
To this end we focus on crossing behavior that is relatively close and directly in front of the AV, as it is relevant for safety and proves the pedestrians trust the technology. \\
We therefore define the following research question:\\
are pedestrians more likely (unnecessarily) to pause or stop and yield the roadway to a driverless vehicle when the vehicle did not signal or indicate to  pedestrians that they had been seen? 
and form the following null hypothesis:\\
\textit{H0: There is no relationship between measured pedestrian crossing behavior and driverless vehicle communication signals. }\\
We performed the field tests in shared spaces in which a traditional safety infrastructure to guide VRU does not exist so that everyone is forced to become more alert and ultimately more cooperative~\cite{jaffe2015}. This scenario is applicable for example in the ``last mile'' with automatic delivery robots.

The remainder of the paper is organized as follows: next section describes related work in the field; section~\ref{sec:fieldtest} details the description of the field test. Section~\ref{sec:algorithmimplementation} describes the  modules that acquire the pedestrian's data which allows a quantitative study of the interaction with the autonomous vehicles; section~\ref{sec:analysis} presents the method to assess the data collected; section~\ref{sec:results} presents the obtained results; and, finally, section~\ref{sec:conclusion} discusses and concludes the work.


\section{Related Work}
\label{sec:relatedwork}

As previously mentioned, communication protocols that make the interaction of driverless vehicles and VRUs possible are necessary to foster trust in the automation~\cite{hussein2016p2v}. 
A lot of literature has been dedicated to the study and interpretation of pedestrian behavior as they interact with vehicles. For example, the authors in  \cite{Schmidt2009} identified the parameters that affect pedestrians at crosswalks in order to predict their intentions, addressing this issue from two distinct perspectives, the pedestrian's and driver's. 

Within this study the authors concluded that interactions were based on a given vehicle's distance rather than Time To Collision (TTC), corroborating the results presented in \cite{Oxley2005}. 
The results from Hamaoka $et. al$ \cite{Hamaoka2013} showed that there are several physical locations on the street where  pedestrians seek to confirm the proximity of a vehicle to ensure their safety. An indicator of this was the frequency of head turning being higher at the edges and middle of crosswalks. Although the previous studies quantitatively established the main parameters governing pedestrian decisions when crossing, they were based on conventional traffic situations with manned vehicles. 
		
In recent studies focusing on the interaction between autonomous vehicles and pedestrians, it was shown that people felt more comfortable crossing the street when a form of response from the side of the vehicle was presented (e.g eye contact) similar to the interaction that occurs with manual driven vehicles \cite{Lagstrom2015} and \cite{Yang2017}.  In the same line of research a survey by the League of American Bicyclists concluded that the inability of pedestrians and cyclists  to communicate and make eye contact with a driverless vehicle increased perceived risk~\cite{leagueofcyclist}. 		
		
The authors in \cite{Matthews} measured the importance of using communication interfaces between the pedestrian and the autonomous vehicle. For their purposes they used a remote-controlled golf cart with an LED word display that explicitly indicated when pedestrians should cross the road in front of them, and they developed a simulator to  test human behavior in this particular situation/setting. Basing their results on a qualitative data collection method, they showed that trust in the technology is dependent on prior knowledge about AV and the distance between both pedestrian and vehicle. 

Further, different early-stage display concepts for interfaces were  evaluated by means of crowdsourcing in~\cite{fridman2017walk} and more advanced interfaces and communication protocols have been tested in \cite{Mahadevan2018} by using, for example, images that follow pedestrians \cite{Chang2017} or implicit forms of communication that included vehicle motion patterns such as breaking \cite{Beggiato2018}. \\
Important groundwork for our line of research was laid in~\cite{demiguel2019}, which identified factors that potentially influenced the perception of a road situation as safe in an environment in which vehicles operated with full driving automation (level 5) in a public space. The analysis of recorded videos and subjective qualitative data established that there were several levels of trust, uncertainty and a certain degree of fear among participants. However, the existence of a communication system to support the interaction with the driverless vehicles was evaluated as positive. 

Although previous studies showed that adding an external monitor or screen to an autonomous vehicle helped VRU to gather relevant information to properly identify the road situation and make the right choices~\cite{demiguel2019}, there are studies such as \cite{Rothenbucher2016} that showed that the patterns in the previous studies  are not decisive in defining a behavior in pedestrians. Moreover, in \cite{Clamann2017} and \cite{Pillai2017} the authors concluded that people's reactions and behavior are determined in greater part by the distance and speed of the vehicle than on the interface presented by the vehicles \cite{Rasouli2019}.

All the previous studies focused on  defining the main factors that influenced pedestrian crossing behavior. However, most part of them relied on qualitative data and in the studies that were based on quantitative data, a Wizard of Oz or OZ paradigm was used to mimic the behavior of the intelligent vehicle. We contribute to the state of the art by presenting in this work quantitative data using an unmanned vehicle in a shared space as explained in the next section.


\section{Field Test Description}
\label{sec:fieldtest}

In order to obtain behavioral patterns of different individuals in the environment, we applied the algorithms described in~\cite{Morales-Alvarez2018} and identified poses adopted by pedestrians in the urban environment when they were exposed to the presence of a driving AV. The environment consisted of a shared space in which segregation of VRUs and vehicles was minimized. In such a scenario, traffic relies more on the informal rules of foot traffic.

The selected scenario was the campus of the University Carlos III in Madrid. The campus contains several green spaces that are connected to the village of Legan\'es and consequently the pedestrians were residents of the area as well as students from the university. 
In this scenario the \gls{iCab} autonomous vehicle (see \cite{Gomez2016}) passed multiple times along a predefined route of 30 meters through a perpendicular flow of pedestrians, creating many opportunities for them to cross in front of the AV. However, the flow of pedestrians could move in multiple directions, such that crossing in front of the AV was not absolutely essential.  

\begin{figure}
	\centering
	\begin{subfigure}[b]{0.48\textwidth}
		\centering
		\includegraphics[width=0.6\textwidth]{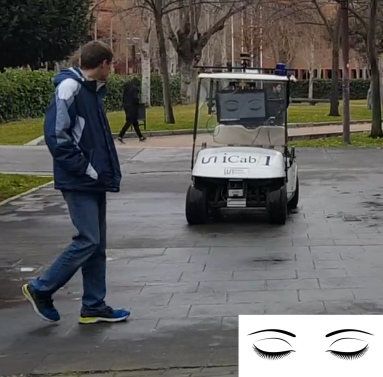}
		\caption{}
		\label{fig:icab:a}
	\end{subfigure}
	\begin{subfigure}[b]{0.48\textwidth}
		\centering
		\includegraphics[width=0.6\textwidth]{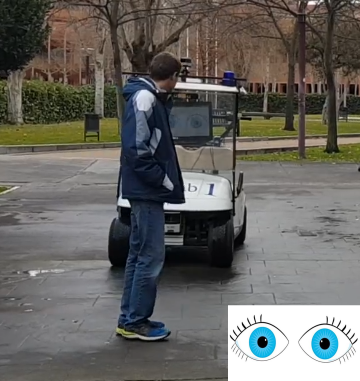}
		\caption{}
		\label{fig:icab:b}
	\end{subfigure}
	\caption{(a) Autonomous vehicle displaying closed eyes to indicate that VRU have not been detected. (b) Open eyes on the display indicating detection of VRU. The image displayed by the AV is depicted on the figure's lower right corner.}
	\label{fig:icab}
\end{figure}

The vehicle was equipped with a external Human Machine Interface (HMI) that conveyed several messages to the pedestrians to indicate whether they had been detected (see Figure~\ref{fig:icab}). During the experiment, vehicle sensors acquired the corresponding data necessary to analyze pedestrian behavior. 

The experiments were conducted for two days where pedestrians were continuously exposed to the \gls{iCab} and the data was recorded for further processing and analysis. From these tests, material of 36 videos was obtained with 135 pedestrians interacting with the vehicle. 
In order to mimic real road conditions as much as possible, pedestrians were not aware of the data collection. To ensure the safety of the pedestrians in case of a failure, a remote control in a fixed location off-site  made it possible to stop the vehicle in an emergency situation.

The following parameters were set to determine road safety as well as response to displayed messages:

\begin{itemize}
\item distance between pedestrians and AV 
\item vehicle's speed along its path
\item head and body pose
\end{itemize} 

Using all of the above data we could obtain pedestrian behavioral patterns indicated by their pose and distance to the AV, as well as road safety-related information such as the TTC (calculated using vehicle speed and pedestrian coordinates). This information was analyzed according to the corresponding image that was displayed on the vehicle interface. 


The design of the HMI relied on the description in~\cite{demiguel2019}. It was developed in C++ and integrated with the Robot Operating System (ROS) into the vehicle's operating system. The vehicle detected pedestrians in the proximity, taking into account the degree of rotation of the vehicle, and then activated different displays depending on whether it had detected the pedestrian or not. 
The algorithms were trained to analyze eye contact, facial expression, and head pose to determine the crossing behavior depending on the message conveyed. To this end the following experiments were performed:

	\subsection{Baseline Condition}

	A performance baseline in which no message was displayed was established to quantify changes in pedestrian behavior.  

	\subsection{Red-Green Sign}

	Inspired by traditional traffic light color-coding, a red screen indicated to pedestrians that it was not safe to cross and a green one signalled that crossing was safe.

	\subsection{Open-Closed Eyes}
	
	An additional set of images mimicked driver behavior as a strategy to ensure that the VRU understood the decisions made by the vehicle. The display showed a pair of open eyes indicating that the pedestrian had been detected and could cross, or a pair of closed eyes indicating that the vehicle had not noticed the pedestrian.


\section{Algorithms Implementation}
\label{sec:algorithmimplementation}

\subsection{Pose Identification}
\begin{figure}
	\centering
	\includegraphics[width=0.48\textwidth]{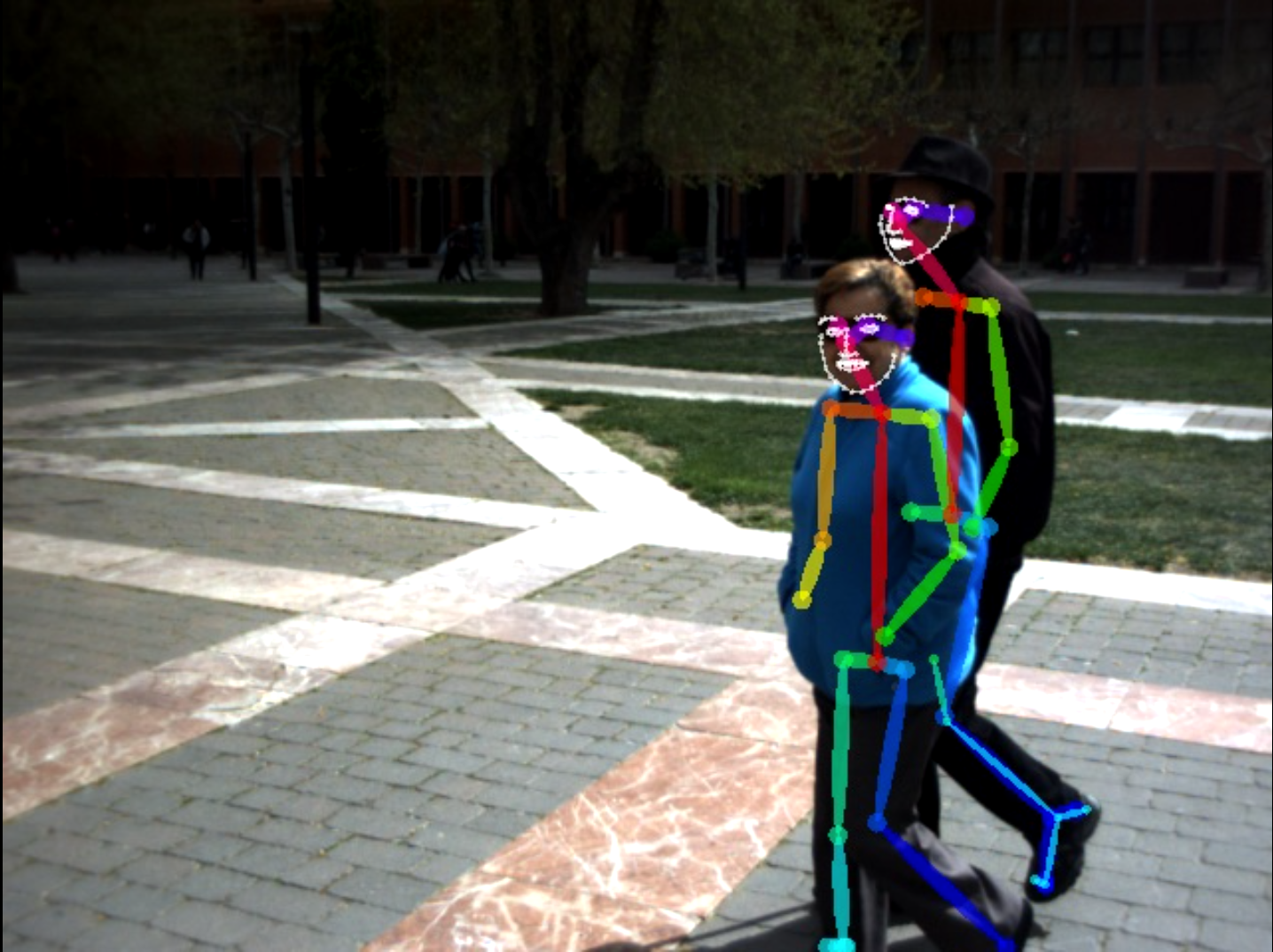}
	\caption{Pose (color points) and face (white points) calculated by OpenPose library.}
	\label{fig:openposerender}
\end{figure}

Relying on the approach presented in \cite{Morales-Alvarez2018}, the specific pose of a pedestrian was identified using the OpenPose open source library developed by CMU-Panoptic labs \cite{cao2017realtime},\cite{cao2018openpose}, \cite{wei2016cpm}, which designed and trained a feedback convolutional neural network that determined key points of individual poses in an RGB image and rendered the poses as seen in Figure~\ref{fig:openposerender}.\\
The neural network is in charge of calculating the heatmaps where the keypoints of the pose are most likely to be found, and it connects them using the \gls{PAFs} feature that preserves the location and orientation of people's joints. 
Using the cameras presented in the vehicle and implementing the OpenPose library, it is possible to obtain up to 25 pedestrian pose keypoints and 26 facial position keypoints to determine behavioral patterns. 

\subsection{Distance Estimation}
\begin{figure}
	\centering
	\includegraphics[width=0.48\textwidth]{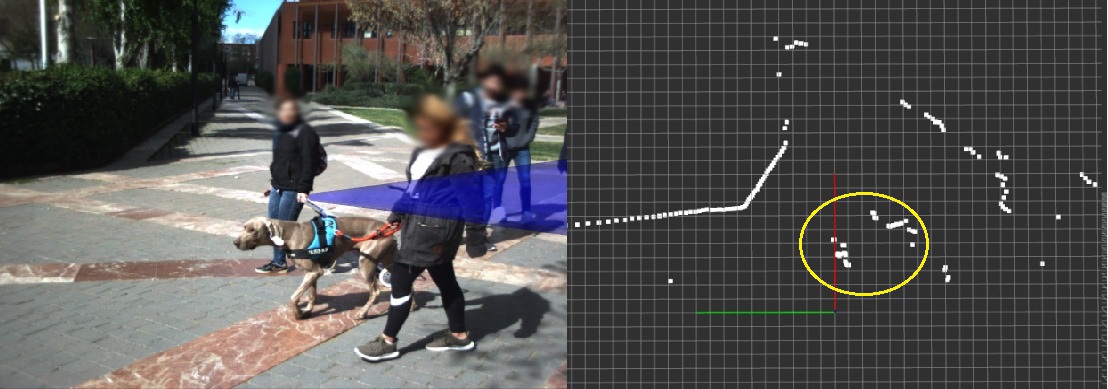}
	\caption{Left: Image acquired by the stereoscopic camera on the autonomous vehicle. The reference axis perpendicular to the picture plane corresponds to the red axis in the image on the right. Right: Representation of VRUs as a series of distance points on the RViz ROS visualization widget. The reference axes are located in the center of the image, whose plane is parallel to the vehicle's plane of movement. }
	\label{fig:laser_scan}
\end{figure}

A further crucial parameter for estimating   road safety is the pedestrian's distance from the approaching vehicle at the time of crossing. Thanks to a 2D laser that is integrated in the autonomous vehicle, it is possible to acquire the distances between the pedestrians and the AV. As in the previous modules, the acquisition of laser's data is obtained through ROS publishing an acquisition node in a certain topic ($/icab/scan$) the distances in meters of the nearby objects. The data is obtained as a series of points that can be observed using the RViz visualization package as shown in Figure \ref{fig:laser_scan}. Using the RVIZ tools and analyzing the points corresponding to pedestrians, it is possible to determine their distance at the time of interaction with the vehicle.

\subsection{Velocity}
The autonomous vehicle was equipped with wheel optical encoders with which the speed of the vehicle could be obtained, taking into account the physical dimensions of the automobile's wheels. The ROS package installed in the \gls{AV} made it possible to publish the speed of the vehicle at all times through the pertinent topic (e.g. $/icab1/velocity\_absolute$).


\section{Data Acquisition and Analysis for Pedestrian Behavior}
\label{sec:analysis}

To test the defined hypothesis we acquired the required data through the algorithms described in section~\ref{sec:algorithmimplementation} as follows:

Head and body poses to determine behavioral patterns depending on the message displayed were identified by the autonomous vehicle. As described in section~\ref{sec:fieldtest} the experiment conditions were as follows: baseline, red-green sign or open-closed eyes. 
Two categories were created based on these data:
\begin{enumerate}
\item VRU that saw the message inside the car and changed their behavior (e.g. stopped for a moment).
			\item VRU that saw the message and continued without any change (e.g. kept walking). 
\end{enumerate}
To test the relationships between the  categorical variables we performed a Pearson ${\chi}^2$ test. Further, we determined the distance between the iCab and the pedestrians. Based on this distance we also calculated the TTC based on the velocity of the vehicle. Statistical significance of the relationship was tested through a unpaired t-test.

The data corresponding to the persons who did not see the vehicle was additionally analyzed. As it is known, eye contact plays a critical role at unmarked intersections, as integrating glances facilitates cooperative action while avoiding eye contact is a way of dominating the other in an interaction~\cite{schelling1984choice} cited in~\cite{vanderbilt2009traffic}. 		
Finally, as in \cite{Olaverri-Monreal2017}, a power analysis was performed to measure the effectiveness of the t-test of rejecting the null hypothesis by calculating the probability of not committing an error of type II (1-$\beta$) or, in other words, the probability of falsely rejecting the null hypothesis.

\section{Results}
\label{sec:results}


From the extracted information we could derive that 92 pedestrians (68.14\%) looked at the screen that was displaying the images and 43 (31.86\%) didn't even look at the vehicle. 

From the pedestrians that looked at the screen it could be observed a greater percentage of pedestrians that crossed in front of the AV, independently of the message displayed. From the results presented in Table~\ref{table:chi:2} the distributions of the categorical variables differed from one another being the differences in the proportion of pedestrians who walked or stopped when the screen showed red or closed eyes not statistically significant. 
Therefore, 
we fail to reject the null hypothesis. 


	\begin{table}
	\centering
	\scriptsize
	\caption{Pedestrian behavior depending on the system display condition}
	\label{table:chi:2}
	\begin{tabular}{|p{1.33cm}|p{1.84cm}|p{0.75cm}|p{0.5cm}|p{0.75cm}|p{0.7cm}|}
		\hline
		&\emph{Baseline}&\emph{Green color}&\emph{Open eyes}&\emph{Red color}&\emph{Closed eyes}\\
		\hline
		Walking&17&9&11&25&21\\
		Stand&3&2&1&1&2\\
		\hline
	\end{tabular}
	\begin{tabular}{|p{8.05cm}|}
		\hline
	    $\chi{}^2$ test ($\alpha$ =0.05)\\
		\hline
	\end{tabular}
	\begin{tabular}{|p{1.33cm}|p{1.84cm}|p{1.68cm}|p{1.89cm}|}
		\hline
		&Baseline vs. Green color&Baseline vs. Open eyes &Baseline vs. Red color\\
		\hline
	\end{tabular}
	\begin{tabular}{|p{1.33cm}|p{0.75cm}|p{0.65cm}|p{0.75cm}|p{0.5cm}|p{0.75cm}|p{0.7cm}|}
		\hline
		&\emph{(1,N=31)}&\emph{p}&\emph{(1,N=32)}&\emph{p}&\emph{(1,N=46)}&\emph{p}\\
		\hline
		&1.99&0.158&0.49&0.484&1.77&0.183\\
		\hline
	\end{tabular}
	\begin{tabular}{|p{1.33cm}|p{1.84cm}|p{1.68cm}|p{1.89cm}|}
		\hline
		&Green color vs. Open eyes& Green color vs. Red color &Red color vs. Closed eyes\\
		\hline
	\end{tabular}
	\begin{tabular}{|p{1.33cm}|p{0.75cm}|p{0.65cm}|p{0.75cm}|p{0.5cm}|p{0.75cm}|p{0.7cm}|}
		\hline
		&\emph{(1,N=23)}&\emph{p}&\emph{(1,N=27)}&\emph{p}&\emph{(1,N=49)}&\emph{p}\\
		\hline
		&0.49&0.484&2.13&0.144&0.50&0.480\\
		\hline
	\end{tabular}
	\begin{tabular}{|p{1.33cm}|p{1.84cm}|p{1.68cm}|p{1.89cm}|}
		\hline
		Open eyes vs. Red color& Open eyes vs. Closed eyes &Baseline vs. Closed eyes&Green color vs. Closed eyes\\ 
		\hline
	\end{tabular}
	\begin{tabular}{|p{0.7cm}|p{0.2cm}|p{0.75cm}|p{0.65cm}|p{0.75cm}|p{0.5cm}|p{0.75cm}|p{0.7cm}|}
		\hline
		\emph{(1,N=38)} & \emph{p} &\emph{(1,N=35)}&\emph{p}&\emph{(1,N=43)}&\emph{p}&\emph{(1,N=34)}&\emph{p}\\
		\hline
		0.33&0.57&0.01&0.974&0.41&0.522&0.65&0.420\\
		\hline
	\end{tabular}
\end{table}

As for the data related to the distance and TTC, results from the power analysis for independent samples from the t-test ranged from 73\% to 97\%, indicating that there is a low probability of having a type II error and erroneously accepting the null hypothesis testing the parameters. Table~\ref{table:differences} shows the obtained values. The analysis indicated that the distance to the vehicle in the moment of crossing was lower under the baseline condition. It also shows that the TTC was lower when the red/green color- coded message and the message showing open/closed eyes was displayed. However these values did not differ significantly between participants.

\begin{table}
	\centering
	\scriptsize
	\caption{Pedestrian distance to the vehicle as well as TTC while crossing depending on the kind of display showed}
	\label{table:differences}
	\begin{tabular}{|p{1.3cm}|p{1.84cm}|p{1.73cm}|p{1.84cm}|}
		\hline
		Metric&Baseline&Red/green color&Opened/closed eyes\\
		\hline
	\end{tabular}
	\begin{tabular}{|p{1.3cm}|p{0.7cm}|p{0.7cm}|p{0.6cm}|p{0.7cm}|p{0.6cm}|p{0.8cm}|}
		\hline
		&\emph{Mean}&\emph{SD}&\emph{Mean}&\emph{SD}&\emph{Mean}&\emph{SD}\\
		\hline
		Distance(m)&6.14&3.56&7.38&3.48&6.88&2.76\\
		TTC (s)&7.31&4.64&4.9&6.23&5.10&7.91\\
		\hline
	\end{tabular}
	\begin{tabular}{|p{8.02cm}|}
		\hline
		T-Test ($\alpha$ =0.05)\\
		\hline
	\end{tabular}
	\begin{tabular}{|p{1.3cm}|p{1.84cm}|p{1.73cm}|p{1.84cm}|}
		\hline
		Metric&Baseline vs. Red/green color&Baseline vs. Opened/Closed eyes&Red/green color vs. Opened/Closed eyes\\
		\hline
	\end{tabular}
	\begin{tabular}{|p{1.3cm}|p{0.7cm}|p{0.7cm}|p{0.6cm}|p{0.7cm}|p{0.6cm}|p{0.8cm}|}
		\hline
		&\emph{t(92)}&\emph{p}&\emph{t(92)}&\emph{p}&\emph{t(92)}&\emph{p}\\
		\hline
		Distance(m)&1.27&0.20&0.85&0.39&0.67&0.55\\
		TTC (s)&1.51&0.13&1.07&0.28&0.90&0.12\\
		\hline
	\end{tabular}
\end{table}

Regarding the persons who did not see the vehicle, results determined by TTC and distance to the vehicle showed that the effect on road safety of the lack of eye contact at unmarked intersections was not significant (Table~\ref{table:notlooking}).  

\begin{table}
	\centering
	\scriptsize
	\caption{Effect of eye contact on interaction with the AV}
	\label{table:notlooking}
	\begin{tabular}{|p{8.08cm}|}
		\hline
		T-test ($\alpha$=0.05)\\
		\hline
	\end{tabular}
	\begin{tabular}{|p{1.6cm}|p{1.5cm}|p{1.54cm}|p{2.14cm}|}
		\hline
		Metric&Without eye contact&Eye contact&T-Test($\alpha = 0.05$)\\
		\hline
	\end{tabular}
	\begin{tabular}{|p{1.6cm}|p{0.7cm}|p{0.37cm}|p{0.7cm}|p{0.4cm}|p{0.7cm}|p{1.0cm}|}
		\hline
		&\emph{Mean}&\emph{SD}&\emph{Mean}&\emph{SD}&\emph{t(133)}&\emph{p}\\
		\hline
		Distance (m)&6.93&3.28&7.81&3.56&1.41&0.1599\\
		TTC (m/s)&5.87&6.71&8.93 &12.22  &1.37&0.1726\\
		\hline
	\end{tabular}
\end{table}

\begin{figure}
	\centering
	\begin{subfigure}[b]{0.48\textwidth}
		\centering
		\includegraphics[width=\textwidth]{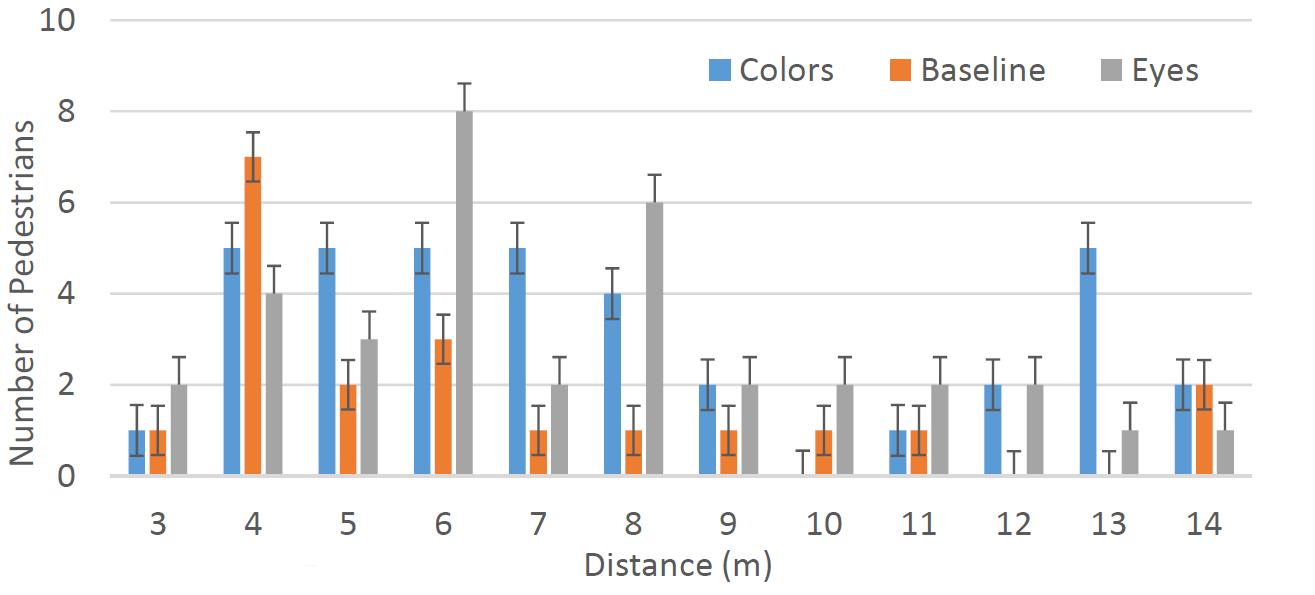}
		\caption{}
		\label{fig:graphics:a}
\end{subfigure}
	\begin{subfigure}[b]{0.48\textwidth}
		\centering
		\includegraphics[width=\textwidth]{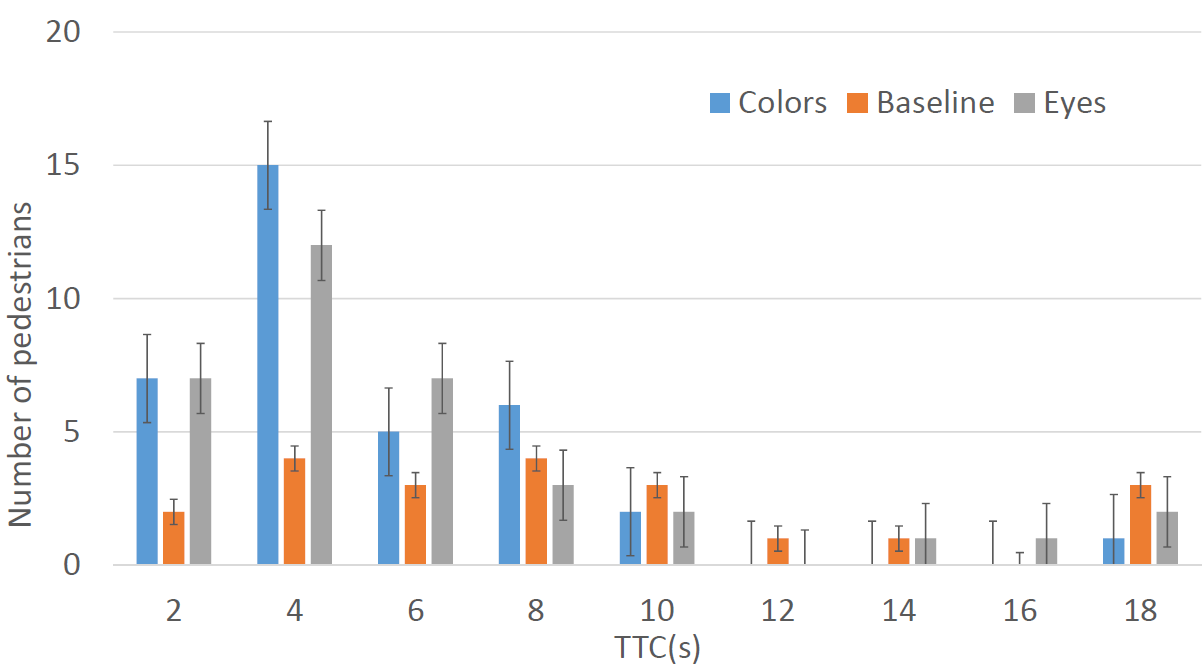}
		\caption{}
		\label{fig:graphics:b}
	\end{subfigure}
	\caption{Pedestrian distance to the vehicle (a) and TTC (b) while crossing depending on display conditions.}
\label{fig:graphics}
\end{figure}

Finally, Figure~\ref{fig:graphics} 
depicts the number of pedestrians that had seen the vehicle and crossed in front of the AV considering their distance to the vehicle, as well as the TTC in relation to the kind of display showed.
From this graphic we can see that 69 pedestrians (71.7\%) crossed at a distance between 5 and 9 meters. The TTC range being 2 to 8 seconds.



\section{Conclusion, Discussion and Future Work}
\label{sec:conclusion}

Initially, it is necessary to note that the fact that looking at the vehicle automatically guaranteed recognition of the display and images on it was confirmed by pedestrian participant comments such as ``the car is looking at you''. This is important because the study is based on the images that the vehicle showed to pedestrians. 

The results reported in this paper did not show statistically significant differences in the proportion of pedestrians who continued walking and crossed in front of the AV to those who stopped depending on the display. Therefore, we fail to reject the null hypothesis.


Moreover, it was observed in most cases that pedestrians crossed even when the message was red or displayed closed eyes. Apparently, the detection of the vehicle on the part of the pedestrians was sufficient to make the decision to cross or not. Therefore, the research question formulated in the beginning:
``Pedestrians are more likely to pause and refrain from crossing in front of an AV'' could not be confirmed.
Furthermore, the kind of display did not affect the distance at which pedestrians crossed in front of the AV and the TTC.
This was probably because the vehicle was slow, never exceeding 5 $m/s$, as people are less likely to respond to a low-speed moving AV,
which is not dangerous to them. As described in the section~\ref{sec:relatedwork} previous works have shown the importance of the vehicle movement (e.g speed, distance) for pedestrians. However they based on simulations or subjective data, while this work describes the quantitative results of a field test performed with a driverless vehicle. 

The relationship between the absence or presence of eye contact on parameters related to road safety such as distance and TTC was not significant. Therefore, it could not be confirmed or disconfirmed whether eye contact with an AV in the tested shared space scenario, in which traffic relies more on the informal rules of foot traffic without traffic lights, road markings or signs that indicate the right-of-way, facilitated cooperative action. \\
During the experiment, it could be observed that in most cases people were distracted, using a cell phone or conversing. For safety reasons, in these cases the AV stopped, causing the pedestrian's curiosity. Interestingly, a high number of pedestrians were first aware of the vehicle only when it stopped.\\
 We can conclude that from the results obtained in section~\ref{sec:results}, the implementation of visual communication cues for interacting with VRUs is not necessarily required for a shared space in which informal traffic rules apply. They are more likely to help when vehicle and pedestrian have potential conflicts that cause danger. 
These results are in line with the findings in \cite{Rothenbucher2016}, \cite{Clamann2017} and \cite{Pillai2017} that stated that information showed on external monitors was not determinant to define a behavior in pedestrians being distance and speed of the vehicle more decisive \cite{Rasouli2019}. 

Therefore, and in line with the finding in~\cite{demiguel2019},  future work will focus on other communication signs such as auditory cues. We will also use additional sensors for cataloging pedestrian behavior that rely on the reconstruction of 3D points to determine, for example, the number of pedestrians who crossed behind the vehicle, as the camera and laser used were only able to record situations that occurred in front of the vehicle.

\section*{ACKNOWLEDGMENT}
This work was supported by the Austrian Ministry for Transport, Innovation and Technology (BMVIT) Endowed Professorship for Sustainable Transport Logistics 4.0.

\bibliographystyle{IEEEtran}
\bibliography{library}
\end{document}